\documentstyle[12pt]{article}

\date{Revised January 1995\\LTP-041-UPR, IIA-NAPP-94-14}
\title{Thermal self-energies at zero momentum}
\author{Jos\'e F. Nieves\\
Department of Physics, University of Puerto Rico\\
P. O. Box 23343, Rio Piedras, Puerto Rico 00931\\
\and
Palash B. Pal\\
Indian Institute of Astrophysics,
Bangalore 560034, INDIA}

\begin{document}

\maketitle

\begin{abstract}

In general the zero momentum limit of thermal self-energies calculated in
perturbation theory depends on the order in which the time and the
space components of the momentum are taken to zero. We show that this
is an artifact of the perturbative calculation, and in fact the
non-analyticity of the one-loop self-energy disappears
when it is calculated with improved vertices
and/or improved propagators
that incorporate the imaginary part of the self-energy.

\end{abstract}

The existing calculations of thermal self-energy functions using
the formalism of Thermal Field Theory
yield results that are not defined if the external
momentum 4-vector is zero.  The classic example is
the photon self-energy $\pi_{\mu\nu}$ in an electron gas. It is well
known \cite{wellknown} that the result of the one-loop calculation
of $\pi_{\mu\nu}(k^0,\vec k)$ for a photon with external momentum
$k^\mu = (k^0,\vec k)$ is such that
\begin{eqnarray}
\lim_{| \vec k | \to 0} \pi_{\mu\nu} (0, \vec k) \neq
\lim_{k^0 \to 0} \pi_{\mu\nu} (k^0, \vec 0)\,,
\label{problem}
\end{eqnarray}
so that the limit in which all components go to zero is not defined.
To be more specific, this inequality is obtained only
for the real part of the self-energy.  For
the imaginary part, the two limits coincide and hence it
is well defined.

There have been two opposite attitudes to this result.
On one side it has been argued
that the inequality should be expected on physical grounds, since
the quantity $\lim_{| \vec k | \to 0} \pi_{\mu\nu}
(0, \vec k)$ describes the screening of static electric fields in the
long wavelength limit, whereas $\lim_{k^0 \to 0} \pi_{\mu\nu}
(k^0, \vec 0)$ gives the
plasma oscillation frequencies in the quasi-static limit, and they
should be different. Others believe that this is a mere restatement
of the inequality in words rather than in symbols, and it does not
explain really why the limit of a static homogeneous field should be
a pathological one. Indeed, Gribosky and Holstein \cite{GrHo90} even
argued, using an effective Lagrangian approach, that the limit of zero
momentum is a perfectly well-defined one. We will side with this
second group and consider the
inequality of Eq.\ (\ref{problem}) as a problem, or a puzzle.

Various attempts have been made to resolve this
puzzle \cite{GrHo90,MNU85,FuYa88,Eva88,BD}, which involve
either introducing new
and ad-hoc Feynman rules for thermal field theories, or putting
restrictions on the general rules \cite{weldon}.  Along another line of
attempt,
it has
recently been pointed out by Arnold, Vokos, Bedaque and Das (AVBD)
\cite{AVBD93} that the problem mentioned above occurs only if the self
energy diagram contains two propagators of the same mass.
If the masses of
the particles in the loop are different, the problem does not exist.
In fact, the calculations of the neutrino self-energy in a gas of
electrons and nucleons, which were carried out even before the work of
Ref.~\cite{AVBD93}, show this  feature  explicitly
\cite{RaNo88,PaPh89,Nie89}.  It has been
speculated that this property may be utilized to introduce
a mass-splitting regularization for thermal diagrams\cite{AVBD93} in
cases where problems are known to occur.

This problem, as well the attempts to resolve it, are
based on the results of one-loop perturbative calculations.
It is natural to ask whether the singular behaviour of the self-energy
function at zero momentum might be a consequence of the approximations
and idealizations that are implicitly made in the perturbative
calculations. In this article we show that this is precisely the case.

We begin our analysis by showing that the problem does
not exist in some theories even if the virtual lines in the loop carry
particles of the same mass.  The example that we give explicitly
in this context is the model of a fermion $\psi$ interacting with
a pseudoscalar $\phi$ via a Yukawa-type interaction of the form
$\overline\psi\gamma_5\psi\phi$.  The 1-loop self-energy of the $\phi$
particle in this model is well defined in the zero momentum limit.
We mention that something similar occurs if the interaction
is taken to be a derivative coupling of the form
$\overline\psi\gamma_\mu\gamma_5\psi\partial^\mu\phi$.
While this result is not too surprising
since the low momentum limit of this theory and the theory
with the pseudoscalar coupling is similar, this example
illustrates in a particularly simple way that the zero
momentum singularity in some cases disappears
if the self-energy diagram
is calculated with an improved vertex.
We then consider some classes of models
in which, as we show, the 1-loop
calculation of the self-energy
yields a result that is defined at zero momentum
when it is carried out using improved propagators
for the particles that appear in the internal lines of the loop
diagrams. Some of these propagators have an absorptive part which, as
observed by AVBD\cite{AVBD93}, is well-defined at zero momentum even
if they are calculated to one-loop and the particles in the internal
lines of the loop have the same mass. As we will see, the
absorptive part of such diagrams in turn
governs the zero-momentum limit of those self-energy diagrams in which
the internal lines have the same mass.
The conclusion that emerges is that, in  theories
where the problem exists, the singular nature
of the 1-loop self-energies in the zero momentum limit
signal a breakdown of the perturbative expansion in that
momentum regime.  In those cases, the calculation must
be carried out using an improved vertex and/or improved propagators
in which the absorptive effects are taken into account.  In a theory
with gauge invariance, such as Scalar Electrodynamics or QED proper,
it is necessary to use both the improved vertex and the improved
charged-particle propagator
since any modification
to either one of these quantities must be accompanied by
a modification to the other, in order to satisfy the Ward
identity and thereby guarantee gauge invariance.

Before proceeding, we recapitulate some results of the
canonical approach to the thermal propagators,
which we will be using
throughout \cite{Nie90}. In this approach, one has to use
anti-time-ordered propagators in addition to the time-ordered
ones, as well as propagators with no time-ordering.
These propagators
can be arranged in the form of a $2\times 2$ matrix. For example, for
any quantum field $\Phi^A$ where $A$ denotes any Lorentz index
carried by the field (none for a scalar field, a Dirac index
for the fermion), we
can write
\begin{eqnarray}
i {\mbox{\boldmath\cal $D$}}_{11}^{AB} (x-y) &\equiv & \phantom{F}
\left< {\cal T} \; \Phi^A (x) \overline\Phi^B (y) \right> \,, \\*
i {\mbox{\boldmath\cal $D$}}_{22}^{AB} (x-y) &\equiv & \phantom{F}
\left< \overline{{\cal T}} \; \Phi^A (x) \overline\Phi^B (y) \right>
\,, \\*
i {\mbox{\boldmath\cal $D$}}_{12}^{AB} (x-y) &\equiv & {\cal F} \left<
\overline\Phi^B (y) \Phi^A (x) \right> \,, \\*
i {\mbox{\boldmath\cal $D$}}_{21}^{AB} (x-y) &\equiv & \phantom{F}
\left< \Phi^A (x) \overline\Phi^B (y) \right> \,,
\end{eqnarray}
where the bar denotes complex conjugation for bosons and Dirac
conjugation for fermions and $\cal F$ is defined by
\begin{eqnarray}
{\cal F} = \left\{ \begin{array}{rl} -1 & \mbox{for fermions,}\\
1 & \mbox{for bosons.} \end{array} \right.
\end{eqnarray}
$\cal T$ and
$\overline{{\cal T}}$ are the time-ordering and anti-time-ordering
operators defined as
\begin{eqnarray}
{\cal T} \; \Phi^A (x) \overline\Phi^B (y)
&\equiv& \Theta(x_0-y_0) \Phi^A (x) \overline\Phi^B (y) + {\cal F}
\Theta (y_0-x_0) \overline\Phi^B (y) \Phi^A (x) \,,\\*
\overline{{\cal T}} \; \Phi^A (x) \overline\Phi^B (y)
&\equiv& \Theta(y_0-x_0) \Phi^A (x) \overline\Phi^B (y) + {\cal F}
\Theta (x_0-y_0) \overline\Phi^B (y) \Phi^A (x) \,,
\end{eqnarray}
where $\Theta$ is the step function.  We  now write the momentum
space expansion of the field in the form
\begin{eqnarray}
\Phi^A (x) = \int {d^3p \over (2\pi)^3 2E} \sum_\lambda \left[
a_\lambda(p) u^A(p,\lambda) e^{-ip \cdot x} +
b^\ast_\lambda(p) v^A(p,\lambda) e^{ip \cdot x} \right] \,,
\end{eqnarray}
where $u^A$ and $v^A$ represent different plane wave solutions
arranged by the index $\lambda$, and $a_\lambda(p)$ and
$b_\lambda(p)$ are the annihilation operators for particles and
antiparticles respectively (for a self-adjoint field
$a_\lambda(p) = b_\lambda(p)$). The properties of  the thermal
bath come in from the expectation values
\begin{eqnarray}
\left< a_\lambda(p) a^\ast_{\lambda'} (p') \right> &=& (2\pi)^3 2E \delta
(\vec p - \vec p\,') \delta_{\lambda\lambda'} \left[1 + {\cal F}f(p, \alpha)
\right] \,, \\*
\left< b_\lambda(p) b^\ast_{\lambda'} (p') \right> &=& (2\pi)^3 2E \delta
(\vec p - \vec p\,') \delta_{\lambda\lambda'} \left[1 + {\cal F}f(p,
-\alpha)\right] \,,
\end{eqnarray}
with
\begin{eqnarray}
f(p,\alpha) = {1 \over e^{\beta p\cdot u - \alpha} - {\cal F}} \,,
\label{f}
\end{eqnarray}
where $\alpha$ plays the role of a chemical potential.
We have introduced the velocity 4-vector $u^\mu$ of the
heat bath, which has components $(1,\vec 0)$ in its own
rest frame.

For future purpose, we introduce the matrix ${\mbox{\boldmath\cal $U$}}$:
\begin{eqnarray}
{\mbox{\boldmath\cal $U$}} = {1 \over \sqrt{1 + {\cal F}
\eta(p,\alpha)}}
\left( \begin{array}{cc} 1 + {\cal F} \eta(p,\alpha) &
\Theta (-p\cdot u) + {\cal F} \eta(p,\alpha) \\
\Theta (p\cdot u) + {\cal F} \eta(p,\alpha) &
1 + {\cal F} \eta(p,\alpha) \end{array} \right) \,,
\end{eqnarray}
where
\begin{eqnarray}
\eta(p,\alpha) = \Theta(p\cdot u) f(p,\alpha) + \Theta(-p\cdot u)
f(-p,-\alpha) \,,
\end{eqnarray}
for both bosons and fermions, provided we use the appropriate form
for the function $f$. In cases where specifically the bosonic or the
fermionic form has to be used, we will use a subscript $B$ or $F$ to
indicate that fact.  Thus,
the procedure described above gives \cite{Nie90} the
$2\times 2$ scalar field thermal propagator as
\begin{eqnarray}
{\mbox{\boldmath $\Delta$}} (p) = {\mbox{\boldmath\cal $U$}}_B
\left( \begin{array}{cc} \Delta_0  & 0 \\ 0 & -\Delta_0^*
\end{array} \right)  {\mbox{\boldmath\cal $U$}}_B \,,
\label{boseprop}
\end{eqnarray}
where
\begin{eqnarray}
\Delta_0 \equiv {1 \over p^2 - M^2 + i0} \,,
\end{eqnarray}
and for fermions
\begin{eqnarray}
{\bf S} (p) =  {\mbox{\boldmath\cal $U$}}_F
\left( \begin{array}{cc} S_0 & 0 \\ 0 & -\widehat S_0 \end{array} \right)
{\mbox{\boldmath\cal $U$}}_F  \,,
\label{fermiprop}
\end{eqnarray}
where
\begin{eqnarray}
S_0 & = & {1 \over \rlap/p - m + i0}\,,\nonumber\\
\widehat S_0 & = &
\gamma_0 S_0^\dagger \gamma_0 \,.
\end{eqnarray}
It is easily checked that this gives, for example,
the propagators
\begin{eqnarray}
\label{S11}
{\bf S}_{11}(p) & = &  (\rlap/p + m_\psi) \left[ {1 \over p^2 - m_\psi^2 +
i0} + 2\pi i \delta (p^2-m_\psi^2) \eta_F(p,\alpha_\psi)\right]\nonumber\\
{\bf \Delta}_{11}(p) & = & \frac{1}{p^2 - M_\phi^2 + i0} -
2\pi i \delta (p^2 - M_\phi^2) \eta_B(p,\alpha_\phi)
\end{eqnarray}
for a fermion $\psi$ and scalar $\phi$,
which are the ones given by Dolan and Jackiw \cite{DoJa74}.
The explicit forms of the other components are also given in the
literature~\cite{others}. We now investigate the implications of
these Feynman rules on self-energy diagrams in various models.

%
%
\subsection*{Model 1}
Consider the pseudoscalar interaction
\begin{equation}
L_{\rm int} = if\overline\psi\gamma_5\psi\Phi + \lambda\Phi^4\,.
\end{equation}
Notice  that the Lagrangian in this case obeys a parity symmetry
   \begin{eqnarray}
\Phi \to -\Phi \,, \quad \psi \to \gamma_0 \psi \,,
\label{Psymm}
   \end{eqnarray}
under which these are the only possible renormalizable
interaction terms. A cubic $\Phi^3$ interaction, for example, is
not invariant under this symmetry.
Using the free-field propagator in Eq.\ (\ref{S11}) for the
fermion field we obtain
\begin{eqnarray}
\label{pinonsingular}
\mbox{Re} \, {\mbox{\boldmath $\Pi$}}^{(\Phi)}_{11}(k_0, \vec k) & = &
4f^2 \int \frac{d^4p}{(2\pi)^3} \eta_F(p,\alpha_\psi)
\delta(p^2 - m_\psi^2)\nonumber\\
& & \quad\times\left\{\frac{p\cdot k}{k^2 + 2p\cdot k}
+ (k\rightarrow -k)\right\}\,,
\end{eqnarray}
which has the unique limit
\begin{eqnarray}
\mbox{Re} \, {\mbox{\boldmath $\Pi$}}^{(\Phi)}_{11}(0, \vec 0) & = &
4f^2 \int \frac{d^3p}{2E(2\pi)^3}(f_\psi + f_{\overline\psi})\,,
\end{eqnarray}
where we have denoted by $f_{\psi,\overline\psi} =
f_F(p,\pm\alpha_\psi)$ the fermion and
antifermion momentum distributions.
In Eq.\ (\ref{pinonsingular})
the vacuum contribution has been omitted,
as we will always do henceforth
whenever we
write explicit expressions for the self-energies.

In passing, we note that the property of a unique
zero-momentum limit does not follow
if the fermion-boson interaction is scalar rather than pseudoscalar.
On the other hand, if the fermion bilinear is either vector or
axial vector type, with a derivative coupling to the spin-0 boson,
the result of the 1-loop diagram gives
$\mbox{Re} \,
{\mbox{\boldmath $\Pi$}}^{(\Phi)}_{11}(0, \vec 0) = 0$
independently of how the limit is taken.  Of course, such couplings
can arise only if the theory at hand is an effective one.
Nevertheless, it shows the possibility that
the zero momentum singularity
goes away
if the self-energy is calculated with an effective vertex,
that  may itself be the result of an improved calculation,
instead of the fundamental one.
In the case of the vector coupling,
the parity transformation of Eq.\ (\ref{Psymm})
is not a symmetry, so
it is likely that the theory includes a trilinear coupling
$\Phi^3$, which gives a contribution to the self-energy that
contains the zero momentum singularity.

\subsection*{Model 2}
Consider a scalar $\Phi$ interacting with another (charged) scalar
$\phi$
with the following interaction Lagrangian
\begin{equation}
\label{model1scalar}
L_{\rm int} = (f\phi\phi\Phi^\ast + H.c.) + \lambda_\phi|\phi|^4
+ \lambda_\Phi|\Phi|^4
\end{equation}
It can be easily seen that these are the most general
renormalizable interaction terms if the Lagrangian obeys a
global U(1) symmetry under which the charges of $\phi$ and
$\Phi$ are 1 and 2 respectively. This symmetry automatically
rules out any cubic self-interaction of any of the scalar
fields.
This model is similar to the following one
\begin{equation}
\label{model1fermion}
L_{\rm int} = (f\psi_L^T C \psi_L\Phi^\ast + H.c.) + \lambda|\Phi|^4
\end{equation}
in which $\Phi$ interacts with a Weyl fermion.

Now consider, for example, the one-loop diagram for the $\Phi$
self-energy in a background
of $\phi$ particles, depicted in Fig.~\ref{f:piPhi}.
The result of calculating that diagram using the free-field
propagator given above for the $\phi$ field is
\begin{eqnarray}
\label{pisingular}
\mbox{Re} \, {\mbox{\boldmath $\Pi$}}^{(\Phi)}_{11}(k_0, \vec k) & = &
2f^2 \int \frac{d^4p}{(2\pi)^3} \eta_B(p,\alpha_\phi) \delta(p^2 - M_\phi^2)
\left(\frac{1}{k^2 - 2p\cdot k}\right)\nonumber\\
& = & 2f^2 \int \frac{d^3p}{(2\pi)^3 2E}\left[\frac{f_\phi}{k^2
- 2p\cdot k}
+ \frac{f_{\overline\phi}}{k^2 + 2p\cdot k}\right]\,,
\end{eqnarray}
where $p^\mu = (E,\vec p)$ with $E = \sqrt{\vec p^2 + M_\phi^2}$
and we have put $f_{\phi,\overline\phi} = f_B(p,\pm\alpha_\phi)$
to denote the momentum distributions.
The above formula reveals the problem to which we alluded
in Eq.\ (\ref{problem}).  The same behaviour is obtained
for the photon self-energy in a background of electrons,
or charged scalars.  The main observation of this paper is that,
as we already mentioned, this problem vanishes if the diagram
is evaluated employing the full propagator of the
$\phi$ field instead of the free-field propagator,
which is what we show now.

In what follows
we will treat in detail
only the two scalar model defined by Eq.\ (\ref{model1scalar}),
since the analysis and results are similar for the other model
of Eq.\ (\ref{model1fermion}).
As we mentioned earlier, this problem vanishes if the diagram
is evaluated employing the full propagator of the
$\phi$ field instead of the free-field propagator.
The full $\phi$ propagator, which we denote
by ${{\mbox{\boldmath $\Delta$}}}^{(\phi)\prime}(p)$,
can be written just like in Eq.\ (\ref{boseprop})
but with $\Delta_0$ replaced by
        \begin{eqnarray}
\Delta^{(\phi)\prime}_0 =  {1 \over p^2 - M_\phi^2 - \Pi^{(\phi)}_0},
        \end{eqnarray}
where $\Pi^{(\phi)}_0$ is the self-energy function for the $\phi$ field.
Thus,
        \begin{eqnarray}
{{\mbox{\boldmath $\Delta$}}}^{(\phi)\prime}(p) = {\mbox{\boldmath\cal $U$}}_B
\left( \begin{array}{cc} \Delta^{(\phi)\prime}_0  & 0 \\ 0 &
-\Delta_0^{(\phi)\prime\ast}
\end{array} \right)  {\mbox{\boldmath\cal $U$}}_B \,,
\label{bosepropexact}
        \end{eqnarray}
and in particular,
        \begin{eqnarray}
{{\mbox{\boldmath $\Delta$}}}^{(\phi)\prime}_{11} (p) =  \frac{1}{p^2 -
M_\phi^2 -
\Pi^{(\phi)}_0}
-2\pi i\rho^{(\phi)}(p)\eta_B(p,\alpha_\phi)\,,
\label{Delta'11}
        \end{eqnarray}
with the spectral density $\rho^{(\phi)}$ given by
\begin{equation}
\label{rho}
\pi\rho^{(\phi)}(p) = -\frac{\mbox{Im}\,\Pi^{(\phi)}_0}
{(p^2 - M_\phi^2 - \mbox{Re}\,\Pi^{(\phi)}_0)^2 +
(\mbox{Im}\,\Pi^{(\phi)}_0)^2}\,.
\end{equation}
As the interactions are turned off, $\rho^{(\phi)}$ approaches
an on-shell delta function.
However, the existence of a non-zero $\mbox{Im}\,\Pi^{(\phi)}_0$
smears the delta function that is present
in Eq.\ (\ref{S11}).  For this
reason it is the easy to see that,
if Diagram~\ref{f:piPhi} is calculated with the propagator
${{\mbox{\boldmath $\Delta$}}}^{(\phi)\prime}_{11}$ for the $\phi$ field
instead of
the free particle propagator,
then the problem of the non-analyticity of
${\mbox{\boldmath $\Pi$}}^{(\Phi)}_{11}(k)$ at zero momentum disappears.
Notice that the dispersive part of $\Pi^{(\phi)}_0$ plays no role
in this argument.  It is not difficult
to see that retaining only the dispersive part of $\Pi^{(\phi)}_0$
and neglecting its absorptive part does not remove
the singularity at zero momentum of ${\mbox{\boldmath
$\Pi$}}^{(\Phi)}_{11}(k)$.

%
%
The next step is to calculate $\Pi^{(\phi)}_0$ and show that in general
it has an absorptive part.
To this end, we recall that the inverse of the full scalar
propagator is given by
        \begin{eqnarray}\label{bospropinv}
{{\mbox{\boldmath $\Delta$}}}'^{-1} (p) = p^2 - M^2 -{\mbox{\boldmath
$\Pi$}}\,,
        \end{eqnarray}
where ${\mbox{\boldmath $\Pi$}}$ is a $2\times 2$ matrix whose components
must be calculated using the Feynman rules of the theory.
Comparing Eqs.\ (\ref{bosepropexact}) and (\ref{bospropinv}), the
following relations are obtained~\cite{Nie90}:
        \begin{eqnarray}
{\mbox{\boldmath $\Pi$}}_{11} &=& \Pi_0 + (\Pi_0 - \Pi_0^\ast )
\eta_B (p,\alpha) \\
{\mbox{\boldmath $\Pi$}}_{22} &=& -\Pi_0^\ast + (\Pi_0 - \Pi_0^\ast )
\eta_B (p,\alpha) \\
{\mbox{\boldmath $\Pi$}}_{12} &=& -(\Pi_0 - \Pi_0^\ast )
\epsilon (p\cdot u) f_B (p,\alpha)
\label{pi12}\\
{\mbox{\boldmath $\Pi$}}_{21} &=& -(\Pi_0 - \Pi_0^\ast )
\epsilon (-p\cdot u) f_B (-p,-\alpha)
\label{pi21}\,.
        \end{eqnarray}
{}From these,  it is easily seen that
        \begin{eqnarray}\label{realparteq}
\mbox{Re}\, \Pi_0 (p) &=& \mbox{Re}\, {\mbox{\boldmath $\Pi$}}_{11}(p) \\
\mbox{Im}\, \Pi_0 (p) &=& {\epsilon(p\cdot u) {\mbox{\boldmath $\Pi$}}_{12}(p)
\over 2i
f_B(p,\alpha)} \,.\label{absparteq}
        \end{eqnarray}
Therefore, to determine $\Pi^{(\phi)}_0$ we must calculate
${\mbox{\boldmath $\Pi$}}^{(\phi)}_{11}$ and
${\mbox{\boldmath $\Pi$}}^{(\phi)}_{12}$, which can be done by evaluating the
diagrams in
Fig.~\ref{f:piphi}. Since the dispersive part $\Pi^{(\phi)}_0$
is not relevant for
resolving the zero momentum problem of the $\Phi$
self-energy, we will not calculate it here. However,
for the consistency of our scheme it is important to stress that
since the internal lines in Fig.~\ref{f:piphi} correspond to particles
of different mass, the function $\mbox{Re}\,{\mbox{\boldmath
$\Pi$}}^{(\phi)}_{11}$
(and hence $\mbox{Re}\,\Pi^{(\phi)}_0$) does not suffer from
the zero momentum problem according
to the observation of AVBD~\cite{AVBD93}.

We now turn the attention to the calculation of the absorptive
part of $\Pi^{(\phi)}_0$.  The simplest way to proceed is to
calculate
${\mbox{\boldmath $\Pi$}}^{(\phi)}_{12}$ and then use Eq.\ (\ref{absparteq}).
Writing
\begin{equation}
k = p^\prime + p \,,
\end{equation}
the application of the Feynman rules to the diagram of
Fig.~\ref{f:piphi} gives
\begin{eqnarray}
\label{eq.piphi1}
-i{\mbox{\boldmath $\Pi$}}^{(\phi)}_{12}(p) =
(if)(-if)\int\frac{d^4p'}{(2\pi)^4}
i{\mbox{\boldmath $\Delta$}}^{(\phi)}_{21}(p^\prime)
i{\mbox{\boldmath $\Delta$}}^{(\Phi)}_{12}(k)\,,
\end{eqnarray}
where the scalar propagators are given by
\begin{eqnarray}
{\mbox{\boldmath $\Delta$}}^{(\Phi)}_{12}(k) & = & -2\pi i\delta(k^2 -
M_\Phi^2)
f_B(k,\alpha_\Phi)\epsilon(k\cdot u)\,,\nonumber\\
{\mbox{\boldmath $\Delta$}}^{(\phi)}_{21}(p^\prime) & = & 2\pi
i\delta(p^{\prime 2} - M_\phi^2)
f_B(-p^\prime,-\alpha_\phi)\epsilon(p^\prime\cdot u)
\end{eqnarray}
Substituting these propagators
into Eq.\ (\ref{eq.piphi1}) and using Eq.\ (\ref{absparteq})
we obtain
\begin{eqnarray}\label{impi0}
\mbox{Im}\, \Pi^{(\phi)}_0(p) & = & (-\pi f^2)\epsilon(p\cdot u)
\int\frac{d^4 p'}{(2\pi)^3}\delta(k^2 - M_\Phi^2)\delta(p^{\prime 2} -
M_\phi^2)
\nonumber\\
& & \times \epsilon(p'\cdot u)\epsilon(k\cdot u)
(f_B(p^\prime,\alpha_\phi) - f_B(k,\alpha_\Phi))\,.
\end{eqnarray}
In writing this formula we have used the identity
\begin{equation}
f_B(-p^\prime,-\alpha_\phi)f_B(k,\alpha_\Phi)
= f_B(p,\alpha_\phi)[f_B(p^\prime,\alpha_\phi) - f_B (k,\alpha_\Phi)]\,,
\end{equation}
which follows from momentum conservation and the fact that
the chemical potentials satisfy
\begin{equation}
\alpha_\Phi = 2\alpha_\phi\,,
\end{equation}
as a consequence of charge conservation.
Eq.\ (\ref{impi0}) can be written in the form
%
%
%
\begin{equation}
\mbox{Im}\, \Pi^{(\phi)}_0 = - \left| p\cdot u \right| \Gamma(p)\,,
\end{equation}
where we have defined
\begin{eqnarray}\label{totalrate}
\Gamma(p) & \equiv & \frac{f^2}{2p\cdot u}
\int{\frac{d^3k}{(2\pi)^3 2\omega}\frac{d^3p'}{(2\pi)^3
2E'}}(2\pi)^4 \nonumber\\
& & \mbox{}\times\left\{\right.\delta^{(4)}(p + p^\prime - k)
[f_\phi(1 + f_\Phi) - f_\Phi(1 + f_\phi)]\nonumber\\
& & \mbox{} + \delta^{(4)}(p - p^\prime - k)
[(1 + f_{\overline\phi})(1 + f_\Phi) - f_{\overline\phi}f_\Phi)]\nonumber\\
& & \mbox{} + \delta^{(4)}(p + p^\prime + k)
[f_\phi f_{\overline\Phi} - (1 + f_\phi)(1 + f_{\overline\Phi})]\nonumber\\
& & \mbox{} + \delta^{(4)}(p - p^\prime + k)
[f_{\overline\Phi}(1 + f_{\overline \phi}) - f_{\overline\phi}(1 +
f_{\overline\Phi})]
\left.\right\}\,.
\end{eqnarray}
For the sake of brevity, we have put
\begin{eqnarray}
f_{\phi,\overline\phi} & = & f_B (p^\prime,\pm\alpha_\phi) \,, \nonumber\\
f_{\Phi,\overline\Phi} & = & f_B (k,\pm\alpha_\Phi) \,,
\end{eqnarray}
to denote the particle and antiparticle distributions,
and in addition,
\begin{eqnarray}
p^{\prime\mu} = (E',\vec p^{\,\prime})\,, &\qquad &
E' = \sqrt{\vec p^{\,\prime 2} + M_\phi^2}\,, \\
k^\mu = (\omega,\vec k)\,, &\qquad & \omega = \sqrt{\vec k^2 + M_\Phi^2}\,.
\end{eqnarray}

The formula for $\Gamma$ given in Eq.\ (\ref{totalrate})
is immediately recognized
as the total rate for a $\phi$ particle of energy $p^0$ and
momentum $\vec p$ (as seen from the rest frame of the medium)
with integrations over the phase space weighted by
the statistical factors appropriate for each process \cite{Wel83}.
Notice that $f^2$ is the
amplitude for the decay processes
$\phi\overline\Phi\rightarrow\overline\phi$ and
$\phi\rightarrow\overline\phi\Phi$,
for the annihilation processes
$\phi\phi\rightarrow\Phi$ and
$\phi\phi\overline\Phi\rightarrow 0$,
as well as for the inverse reactions of all of them.
For certain specific values
of $p^0$ and $\vec p$ some of these processes will
be kinematically forbidden, but in general $\Gamma$ is non-zero.

%
%
In conclusion, we have presented various models
that have similar particle
content but different interactions,
which exemplify several situations.
In some models the one-loop
self-energies are well defined at zero momentum while in others
the zero momentum singularity
dissapears if the coupling is taken to be an effective
vertex with a suitable momentum dependence.
In another type of model, the singularity
disappears if the self-energy is calculated
with improved propagators that include the absorptive
part.
Carrying over this idea to the case of QED,
which would require that we also use an improved
vertex in order to maintain the Ward identity and
guarantee a gauge invariant result, it implies that
the one-loop photon self-energy calculated with the full
charged particle propagator
instead of the free propagator
is defined at zero momentum.  In particular, the absorptive
part of the charged particle self-energy, which physically
is related to the damping rate of the particle,
cannot be neglected if the photon self-energy
is evaluated at zero momentum.  Then, the physical
picture that emerges is the following.
The traditional formulas that are given for
\begin{eqnarray}
\lim_{| \vec k | \to 0} \pi_{\mu\nu} (0, \vec k)
\end{eqnarray}
and
\begin{eqnarray}
\lim_{k^0 \to 0} \pi_{\mu\nu} (k^0, \vec 0)\,,
\end{eqnarray}
which are related to well known physical quantities
such as the plasma frequency and Debye radius,
are valid in the limiting cases
\begin{eqnarray}
k^0 = 0\,&;& \Gamma \ll| \vec k| \ll m \\
\vec k = 0\,&;& \Gamma \ll k^0 \ll m\,,
\end{eqnarray}
where $m$ stands for the charged particle mass.
Since the two limits correspond to two different
physical situations the results are different.
Traditionally $\Gamma$ is omitted in the above conditions,
but then it must be kept in mind that the formulas
cannot be taken literally all the way to zero momentum.

\newpage

\newpage

\setlength{\unitlength}{1.5mm}

				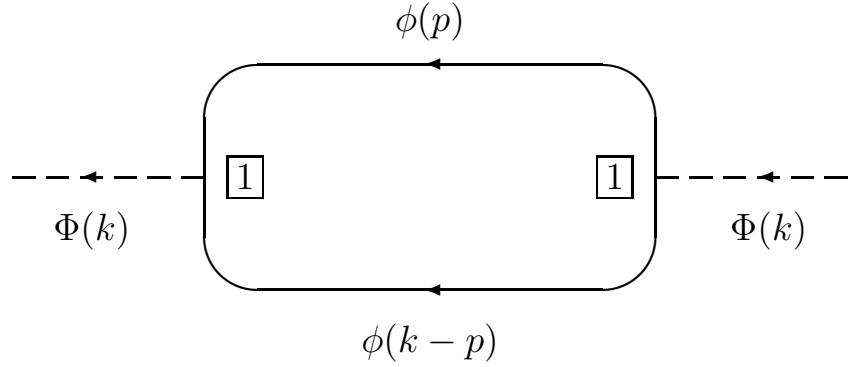
\begin{figure}
		\begin{center}
\large
	\begin{picture}(60,40)(10,10)
\thicklines
\put(40,30){\oval(40,20)}
\put(40,40){\makebox(0,0){\vector(-1,0){1}}}
\put(40,43){\makebox(0,0)[b]{$\phi(p)$}}
\put(40,20){\makebox(0,0){\vector(-1,0){1}}}
\put(40,17){\makebox(0,0)[t]{$\phi(k-p)$}}
\multiput(60,30)(3,0){6}{\line(1,0){2}}
\put(70,30){\vector(-1,0){1}}
\put(70,27){\makebox(0,0)[t]{$\Phi(k)$}}
\multiput(20,30)(-3,0){6}{\line(-1,0){2}}
\put(10,30){\vector(-1,0){1}}
\put(10,27){\makebox(0,0)[t]{$\Phi(k)$}}
\put(22,30){\makebox(0,0)[l]{\fbox{1}}}
\put(58,30){\makebox(0,0)[r]{\fbox{1}}}
	\end{picture}
		\end{center}
\caption[]{\small\sf
Self-energy diagram for a scalar
$\Phi$ in a background of $\phi$ scalar particles
in the model of Eq.\ (\ref{model1scalar}).
The type of each vertex used is depicted in a box near the
corresponding vertex.}\label{f:piPhi}
				\end{figure}

				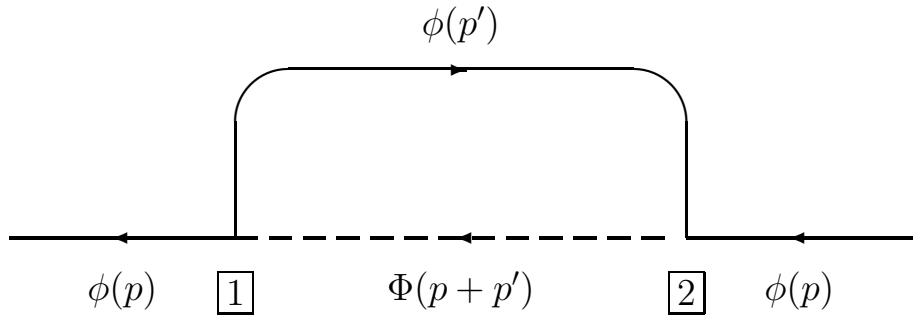
\begin{figure}
		\begin{center}
\large
	\begin{picture}(80,40)(0,20)
\thicklines
\put(40,30){\oval(40,30)[t]}
\put(40,46){\makebox(0,0){\vector(1,0){1}}}
\put(40,48){\makebox(0,0)[b]{$\phi(p')$}}
\put(60,30){\line(1,0){20}}
\put(70,30){\vector(-1,0){1}}
\put(70,27){\makebox(0,0)[t]{$\phi(p)$}}
\put(20,30){\line(-1,0){20}}
\put(10,30){\vector(-1,0){1}}
\put(10,27){\makebox(0,0)[t]{$\phi(p)$}}
\multiput(20,30)(3,0){13}{\line(1,0){2}}
\put(40,30){\makebox(0,0){\vector(-1,0){1}}}
\put(40,27){\makebox(0,0)[t]{$\Phi(p+p')$}}
\put(20,27){\makebox(0,0)[t]{\fbox{1}}}
\put(60,27){\makebox(0,0)[t]{\fbox{2}}}
	\end{picture}
		\end{center}
\caption[]{\small\sf
 Self-energy diagram for $\phi$ in the model of
Eq.\ (\ref{model1scalar}).  For the calculation
of $\mbox{\boldmath $\Pi$}^{(\phi)}_{12}$,
the left and right vertices should be of type 1 and 2,
respectively.}\label{f:piphi}
				\end{figure}


\begin{thebibliography}{[00]}

\bibitem{wellknown} See, e.g., H. A. Weldon, Phys. Rev. D26, 1394, (1982).

\bibitem{GrHo90} P. S. Gribosky, B. R. Holstein, Z. Phys. C47, 205 (1990).

\bibitem{MNU85} H. Matsumoto, Y. Nakano, H. Umezawa, Phys. Rev. D31, 1495
(1985).

\bibitem{FuYa88} Y. Fujimoto, H. Yamada, Z. Phys. C37, 265 (1988).

\bibitem{Eva88} T. S. Evans, Z. Phys. C41, 333 (1988).

\bibitem{BD} P. F. Bedaque and A. Das, Phys. Rev. D45, 2906 (1992).

\bibitem{weldon}  See H. A. Weldon, Phys. Rev. D47, 594 (1993),
for an enlightening discussion of the issues and subtleties
involved.

\bibitem{AVBD93} P. Arnold, S. Vokos, P. Bedaque, A. Das, Phys. Rev.
D47, 4698 (1993).

\bibitem{RaNo88} D. N\"{o}tzold and G. Raffelt, Nucl. Phys. B307, 924
(1988).

\bibitem{PaPh89} P. B. Pal and T. N. Pham, Phys. Rev. D40, 714 (1989).

\bibitem{Nie89} J. F. Nieves, Phys. Rev. D40, 866 (1989).

\bibitem{Nie90} J. F. Nieves, Phys. Rev. D42, 4123 (1990).

\bibitem{DoJa74} L. Dolan, R. Jackiw, Phys. Rev. D9, 3320 (1974).

\bibitem{others} See, e.g., Ref. \cite{Nie90} and references therein.

\bibitem{Wel83} This result provides
        another example of the kind of relations discussed in
        H. A. Weldon, Phys. Rev. D28, 2007 (1983).


\end{thebibliography}
\end{document}